\documentclass[aps,prd,tightenlines,twocolumn,nofootinbib,preprintnumbers,amssymb,
superscriptaddress,showpacs]{revtex4}

\usepackage{amsmath}
\usepackage{graphicx}
\usepackage{epsfig}
\usepackage{multirow}
\usepackage{dcolumn}
\usepackage{bm}
\usepackage[usenames]{color}

\usepackage{epsfig}

\newcommand{\newc}{\newcommand}

\newc{\nn}{\noindent}
\newc{\non}{\nonumber}

\newcommand{\bea}{\begin{eqnarray}}
\newcommand{\eea}{\end{eqnarray}}
\def\beq{\begin{equation}}
\def\eeq{\end{equation}}
\def\baz{\begin{array}{cc}}
\def\ea{\end{array}}

\def\3e{e \bar{e} e}
\def\m2eg{\mu \to e \gamma}
\def\m2{\mu \to}
\def\m2e{\mu \to e}
\def\t2eg{\tau \to e \gamma}
\def\t2mg{\tau \to \mu \gamma}
\def\gs{\mathrel{
   \rlap{\raise 0.511ex \hbox{$>$}}{\lower 0.511ex \hbox{$\sim$}}}}
\def\ls{\mathrel{
   \rlap{\raise 0.511ex \hbox{$<$}}{\lower 0.511ex \hbox{$\sim$}}}}

\bf

\begin{document}
\mbox{ }\\[2cm]
\title{Lower Limits on $\mu \to e \gamma$ from new Measurements on $U_{e3}$}
\author{Joydeep Chakrabortty}%
\email{joydeep@prl.res.in}
\affiliation{Physical Research Laboratory (PRL), 
Navrangpura, Ahmedabad 380009, Gujarat, India}
\author{Pradipta Ghosh}%
\email{pradipta.ghosh@uam.es}
\affiliation{Departamento de F\'{\i}sica Te\'{o}rica UAM and
Instituto de F\'{\i}sica Te\'{o}rica UAM/CSIC, \\
Universidad Aut\'{o}noma de Madrid (UAM), Cantoblanco, 28049 Madrid, Spain
}
\author{Werner Rodejohann}%
\email{werner.rodejohann@mpi-hd.mpg.de}
\affiliation{Max--Planck--Institut f\"ur Kernphysik,
Saupfercheckweg 1, 69117 Heidelberg, Germany}
%
\begin{abstract}
%
New data on the lepton mixing angle $\theta_{13}$ imply that the
$e\mu$ element of the matrix $m_\nu m_\nu^\dagger$, where $m_\nu$ is
the neutrino Majorana mass matrix, cannot vanish.  This implies a lower
limit on lepton flavor violating processes in the $e\mu$ sector in a variety of
frameworks, including Higgs triplet models or the concept of minimal
flavor violation in the lepton sector.  We illustrate this for the 
branching ratio of $\mu \to e \gamma$ in the
type II seesaw mechanism, in which a Higgs triplet is responsible for
neutrino mass and also mediates lepton flavor violation. 
We also discuss processes like $\mu\to e\bar{e}e$
and $\mu\to e$ conversion in nuclei. Since these processes have
sensitivity on the individual entries of $m_\nu$, their rates can
still be vanishingly small. 
\end{abstract}

\preprint{FTUAM-12-88, IFT-UAM/CSIC-12-29, April 2012}


\maketitle

\section{\label{sec:intro}Introduction}
The observation of lepton mixing in the form of neutrino oscillations 
shows without doubt that there is physics beyond the Standard Model of
elementary particles. To be precise, the presence of lepton flavor
violation (LFV) has been established. While being well-entrenched in
the neutrino sector, the question arises how large LFV in the charged 
lepton sector is, and how it is connected to the quantities in the 
neutrino sector. The power of the Glashow-Iliopoulos-Maiani  
mechanism \cite{Glashow:1970gm} in the Standard Model ensures that for
instance observation of $\mu \to e \gamma$ will  
be unambiguously a sign of new physics beyond the
presence of ``only'' massive neutrinos. If this new 
physics is connected to neutrino mixing parameters is an extremely 
model-dependent question.  

In this short note we point out an interesting new implication for
scenarios in which LFV is governed by $m_\nu m_\nu^\dagger$, where
$m_\nu$ is the neutrino mass matrix. In particular the $e\mu$ entry of
this matrix is of interest, as it is often responsible for   
$\mu \to e \gamma$, $\mu \to 3e$, or muon-to-electron conversion in
nuclei. The advantage of scenarios in which $m_\nu m_\nu^\dagger$
governs LFV is their predictivity: $m_\nu m_\nu^\dagger$ only depends on measurable
neutrino oscillation parameters: both mass-squared differences
including the sign of the atmospheric one, three mixing angles and the
Dirac CP phase.    
Until very recently neutrino data allowed for the possibility that
$(m_\nu m^\dagger_\nu)_{e\mu}$ vanishes, namely when the lepton mixing
matrix element $|U_{e3}|$ takes a small value around 0.015. However,
recent results from T2K \cite{Abe:2011sj}, Double Chooz
\cite{Abe:2011fz} and finally Daya Bay \cite{An:2012eh}\footnote{After
completion of the paper, the RENO collaboration reported a new
measurement \cite{reno}, resulting in $|U_{e3}| =
0.163^{+0.013}_{-0.014}$ at $1\sigma$. Our results hardly change by
considering this range of values.} imply a
surprisingly large value of the lepton mixing matrix element
$|U_{e3}|$ around 0.15: 
\beq \label{eq:DB}
|U_{e3}| = 0.153^{+0.014\,(0.039)}_{-0.015\,(0.055)} , 
\eeq
where we have given the $1\sigma$ and $3\sigma$ ranges. As we will
see, this sizable value implies that $(m_\nu m^\dagger_\nu)_{e\mu}$ cannot vanish,
and hence a lower limit on $(m_\nu m^\dagger_\nu)_{e\mu}$ arises.
Correspondingly, lower limits on lepton flavor violating processes
arise. Of course, the processes can still be unobservable because 
of too heavy masses of the additional particles which mediate
the decays.  However, the point here is that the  flavor physics part
of the problem cannot spoil observation anymore. Thereby, yet another
possibility for LFV to hide from future experiments is ruled out.    

A popular example for which the rates of LFV processes are functions of $m_\nu
m_\nu^\dagger$ is the type II (or triplet) seesaw mechanism 
\cite{Konetschny:1977bn,Mohapatra:1979ia,Magg:1980ut,
Lazarides:1980nt,Schechter:1980gr,Cheng:1980qt}. Here neutrino 
mass is generated by a Higgs triplet, which in turn can mediate LFV,
and in particular leads to a branching ratio of $\mu \to e \gamma$ 
depending on $(m_\nu m^\dagger_\nu)_{e\mu}$. 
We focus here on the triplet seesaw mechanism, but point out that 
$m_\nu m_\nu^\dagger$ governs LFV also in classes of theories in which
``minimal flavor violation" in the lepton sector is realized
\cite{mfv}. Minimal flavor violation assumes that Standard Model
Yukawa couplings are the only sources of flavor symmetry
breaking. This very economical and elegant concept was originally
invented for the quark sector \cite{mfv_quark}, but can be applied to
the lepton sector as well \cite{mfv}, predictions for LFV rates depending
however on the explicit operator realization.  Also for the supersymmetric triplet
seesaw, with a very heavy triplet and universal boundary conditions \cite{Rossi:2002zb},
consequences of our observation arise, absolute rates depending however on a variety
of additional parameters. Another explicit 
realization of Br$(\mu \to e \gamma)=f[(m_\nu m^\dagger_\nu)_{e\mu}]$ can be found in 
\cite{Davidson:2009ha}; here neutrinos are Dirac particles within
a particular two Higgs Doublet Model. There are presumably many more
examples. For definiteness, we consider here only the triplet seesaw,
where there are only two free parameters besides the ones governing neutrino
oscillations, namely the mass of the triplet and the vacuum
expectation value of its neutral component.  

The same result for $|U_{e3}|$ implies that $(m_\nu m^\dagger_\nu)_{e\tau}$ 
cannot vanish anymore, and lower limits on $\tau e$ LFV processes arise. 
However, due to the approximate $\mu$--$\tau$ symmetry of lepton mixing, it holds that 
$(m_\nu m^\dagger_\nu)_{e\tau}\sim (m_\nu
m^\dagger_\nu)_{e\mu}$. This implies that rates for $\tau e$ LFV processes
are of the same order as rates for $\mu e$ LFV processes. 
Since future limits on the $\tau  e$ sector are 
expected to be less stringent than present
constraints on the $\mu  e$ sector, those decay channel are not
observable in this framework. This in turn implies that for instance
observation of $\tau \to e \gamma$ will signal the presence of lepton
flavor violation not depending on $m_\nu m_\nu^\dagger$. 

The processes as $\mu \to 3e$ and $\mu-e$ conversion have some
dependence on $(m_\nu m_\nu^\dagger)_{e\mu}$ as well. 
However, either the contribution of $(m_\nu
m_\nu^\dagger)_{e\mu}$ is suppressed, or cancellations from other
contributions can occur. Setting lower limits in the same sense as for
$\mu \to e\gamma$ is not possible. 

The paper is build up as follows: in Section \ref{sec:obs} we quantify
the fact that new oscillation data for large $|U_{e3}|$ imply
non-vanishing $(m_\nu m^\dagger_\nu)_{e\mu}$. Section \ref{sec:main}
introduces the type II seesaw and relevant expressions for lepton
flavor violating processes. A numerical study of the various
constraints is performed in Section \ref{sec:num}, before we conclude
in Section \ref{sec:concl}.

\section{\label{sec:obs}Non-vanishing $|U_{e3}|$ 
and non-vanishing $(m_\nu m^\dagger_\nu)_{e\mu}$}
In this section we note the simple yet consequential fact
that large $|U_{e3}|$ implies non-vanishing $(m_\nu
m^\dagger_\nu)_{e\mu}$. As stated in the introduction, a variety of
scenarios and frameworks leads to LFV processes depending on the
quantity $m_\nu m^\dagger_\nu$. Here  $m_\nu$ is the neutrino mass
matrix which is given as
\beq\label{eq:mnu-mat}
m_\nu = U \, {\rm diag}(m_1, m_2, m_3) \, U^T , 
\eeq
where $m_i$ are the three light neutrino masses and $U$ 
the  Pontecorvo-Maki-Nakagawa-Sakata (PMNS) lepton mixing 
matrix. Its standard parametrization is  
\begin{widetext}
 
\bea\label{eq:U}
U &=&
\left(\begin{array}{ccc}
{c_{12}}{c_{13}} & {s_{12}}{c_{13}} & {s_{13}}  e^{-i\delta}\\ \\
-{s_{12}}{c_{23}}-{c_{12}}{s_{23}}{s_{13}} e^{i\delta} & {c_{12}}{c_{23}}
-{s_{12}}{s_{23}}{s_{13}} e^{i\delta} & {s_{23}}{c_{13}}\\ \\
{s_{12}}{s_{23}}-{c_{12}}{c_{23}}{s_{13}} e^{i\delta} & -{c_{12}}{s_{23}}
-{s_{12}}{c_{23}}{s_{13}} e^{i\delta} & {c_{23}}{c_{13}}
\end{array}\right)
\times{\rm{diag}}
\left(\begin{array}{ccc}
 1,  e^{i\phi_1}, e^{i\phi_2}
\end{array}\right).
\eea
\end{widetext}
In Eq.~(\ref{eq:U}), $\delta$ denotes the Dirac CP-phase, while
$\phi_1,\phi_2$ denote two Majorana phases. The quantities $c_{ij}$
and $s_{ij}$ represent $\cos \theta_{ij}$ and $\sin \theta_{ij}$,
respectively.

\begin{center}
\begin{figure*}[t]
\includegraphics[width=17.65cm,height=4.70cm,angle=0]{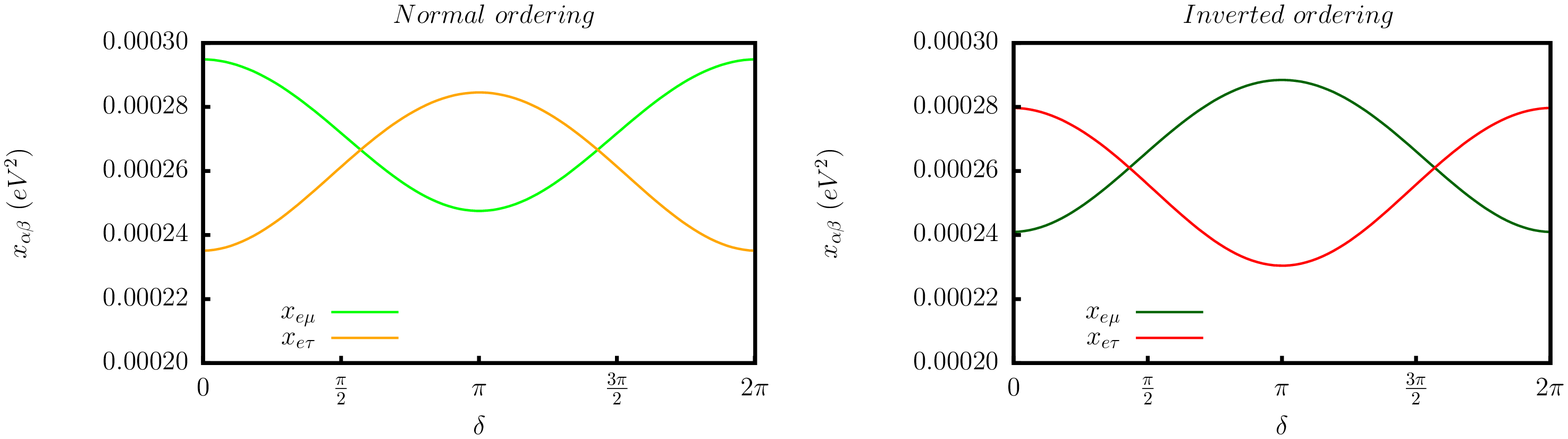} \\[.5cm]
\includegraphics[width=17.65cm,height=4.70cm,angle=0]{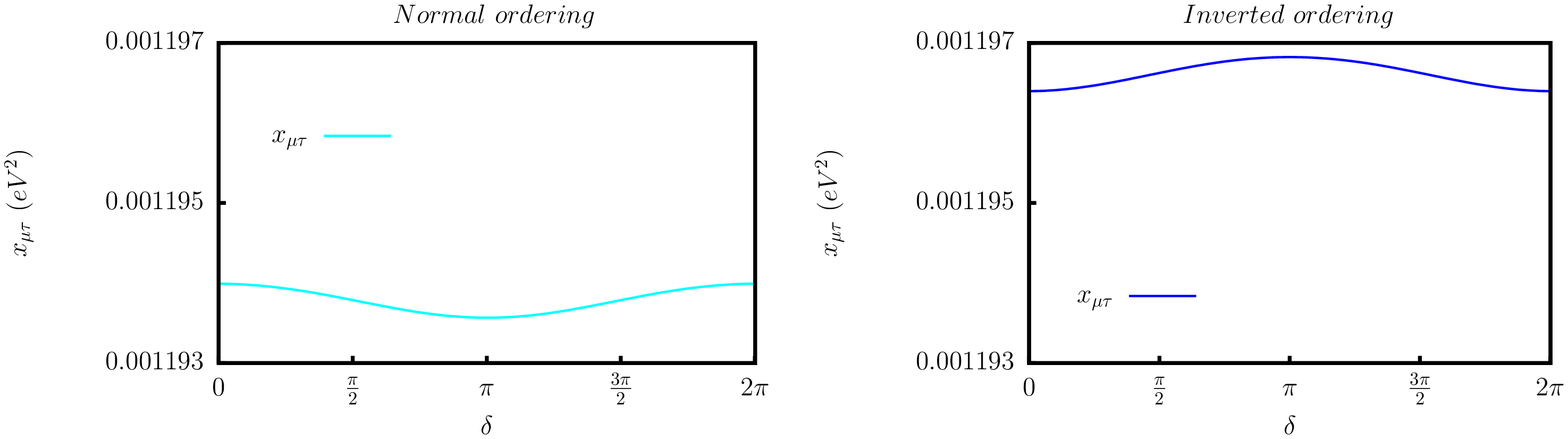}
\caption{ \small {Plots showing the effect of the Dirac CP-phase
$\delta$ on various $x_{\alpha\beta}$ where $x\equiv|(m_\nu m_\nu^\dagger)|$. The remaining
oscillation parameters are fixed at their best-fit values (see 
Table~\ref{tab:osc}).}} 
\label{f:mnumnudagij}
\end{figure*}
\end{center}

We consider here classes of theories in which LFV is governed by $m_\nu
m_\nu^\dagger$. Note that this matrix is independent of the Majorana
phases and the interesting off-diagonal entries are furthermore
independent of the neutrino mass scale (recall that $m_\nu
m_\nu^\dagger$ is the same quantity which appears in the classical
Hamiltonian for neutrino oscillations). We plot in
Fig.~\ref{f:mnumnudagij} the $\delta$ dependency of the off-diagonal
elements of $m_\nu m_\nu^\dagger$, fixing the remaining parameters to
their best-fit values. 
It is apparent (and well-known) that $|(m_\nu m_\nu^\dagger)_{\mu\tau}|$ is larger than
the other entries by one order of magnitude, that $|(m_\nu
m_\nu^\dagger)_{e\mu}|   \sim |(m_\nu m_\nu^\dagger)_{e\tau}|$, and
that the variation of $|(m_\nu m_\nu^\dagger)_{\mu\tau}|$
with $\delta$ is much smaller compared to that of the other two 
off-diagonal entries. Such studied have been performed several times
in the literature before \cite{Rossi:2002zb,prev_LFV,prev_LHC} and 
also recently \cite{Dinh:2012bp}, and here we wish 
to focus on the implication of non-vanishing and sizable $|U_{e3}|$ on
$|(m_\nu m_\nu^\dagger)_{e\mu}|$ and thus on $\mu \to e \gamma$.

In those cases, in which LFV depends on $m_\nu 
m_\nu^\dagger$, the $e\mu$ entry is of particular interest, as in the
$e\mu$ sector the strongest experimental limits on LFV exist, and even
stronger limits are to be expected in the near future
\cite{Hoecker:2012nu,future-LFV}.  
The crucial flavor physics quantity is therefore $(m_\nu 
m_\nu^\dagger)_{e\mu}$. 
One might therefore wonder whether $(m_\nu 
m_\nu^\dagger)_{e\mu}$ can vanish in principle. This is indeed
possible, and the result for $(m_\nu m_\nu^\dagger)_{e\mu} = 0$ is a
rather simple formula: 
{\footnotesize
\bea \label{eq:12zero}
|U_{e3}|_{(m_\nu  m_\nu^\dagger)_{e\mu}=0} &&= \frac{1}{2} \, \frac{R \, \sin 2 \theta_{12} \, 
\cot \theta_{23} }{1 \mp R \, \sin^2 \theta_{12}} 
\simeq \frac{1}{2} \, R \, \sin 2 \theta_{12} 
\cot \theta_{23}  \nonumber\\
&&= \left\{ \baz 
0.0135_{-0.002\,(0.005)}^{+0.004\, (0.009)}  & \mbox{normal}\, , \\[.2cm]
0.0141_{-0.003\,(0.005)}^{+0.003\, (0.009)}  & \mbox{inverted} ,
\ea \right.
\eea}
where the minus (plus) sign is for the normal (inverted) mass
ordering and $R$ is the positive ratio of the solar and the atmospheric mass-squared
differences ($\Delta m^2_{sol}$ and $\Delta m^2_{atm}$, 
respectively)\footnote{Interestingly, the above
condition on $|U_{e3}|$ requires in addition CP conservation,
i.e.~$\delta= 0$ and $\pi$, respectively. Note that with $f = m_\nu 
m_\nu^\dagger$ the Jarlskog invariant for leptonic CP violation in neutrino
oscillations is proportional to Im$(f_{e\mu} f_{e\tau} f_{\mu\tau})$
\cite{jcp}. Hence, the vanishing of an off-diagonal element of $m_\nu 
m_\nu^\dagger$ implies CP conservation.}. We have also given the
implied value of $|U_{e3}|$ when the best-fit values as well as $1\sigma$ and 
$3\sigma$ ranges of the oscillation parameters from Ref.~\cite{Schwetz:2011zk} are inserted.  
The value of $|U_{e3}|$ for which  $(m_\nu  m_\nu^\dagger)_{e\mu} $
vanishes is rather small, being of order $ 0.014$. It has to be
compared to the value of $|U_{e3}| = 0.153^{+0.039}_{-0.055} $
determined by Daya Bay, given in Eq.~(\ref{eq:DB}), which is
significantly larger. This implies a non-zero lower limit on $(m_\nu
m_\nu^\dagger)_{e\mu} $, and hence on branching ratios for lepton
flavor violating processes in a variety of scenarios. This is the main
point of this paper, and we will quantify this for the example of
Higgs triplets in the type II seesaw mechanism. 
Values of the oscillation parameters in the $1\sigma$ and $3\sigma$
range are given in Table~\ref{tab:osc}.  Using those values, the
explicit range at $1\sigma$ and $3\sigma$ of $(m_\nu
m_\nu^\dagger)_{e\mu} $ reads: 
\beq
|(m_\nu  m_\nu^\dagger)_{e\mu} | ~ [{\rm eV}^2] = 
\left\{ \baz 
  1.9 \times 10^{-4} - 4.5 \times 10^{-4} & (1\sigma) , \\
  1.0 \times 10^{-4} - 3.5 \times 10^{-4} & (3\sigma) , 
\ea
\right. 
\eeq
with differences between the normal and inverted ordering not showing
up before the second decimal place. 
Using the recent RENO
result \cite{reno} would give minimal (maximal) values smaller
(larger) by about $0.2 \times 10^{-4}$ eV$^2$. 
\begin{center}
\centering
\begin{table}[t]
\footnotesize
\caption{\label{tab:osc}
best-fit, $1\sigma$ and $3\sigma$ ranges of the oscillation
parameters. Values of all the parameters (except $\sin^2 \theta_{13}$)
are taken from Ref.~\cite{Schwetz:2011zk}. For $\theta_{13}$, the results
of Daya Bay~\cite{An:2012eh} have been used. Results applying for the
inverted mass ordering are in square brackets. 
}
\begin{ruledtabular}
\begin{tabular}{cc c c c}
& Parameters &best-fit& $1\sigma$ range & $3\sigma$ range\\ \hline
& $\Delta m^2_{sol} ~ [{\rm eV^2}]\times10^5$ &7.59& $7.41-7.79$ & $7.09-8.19$\\
& $\Delta m^2_{atm} ~ [{\rm eV^2}] \times10^3$&2.50& $2.34-2.59$&
$2.14-2.76$ \\
& &[-2.40]& $[-(2.31-2.48)]$& $[-(2.13-2.76)]$ \\ 
& $\sin^2 \theta_{12}$ &0.312& $0.297-0.329$ & $0.27-0.36$ \\
& $\sin^2 \theta_{23}$&0.52&$0.45-0.58$ & $0.39-0.64$ \\
&  &[0.52]&$ [0.46-0.58]$ & $[0.39-0.64]$ \\
& $\sin^2 \theta_{13}$ & $0.0236$ & $0.0190-0.0279$ & $0.0097-0.0369$  \\
& $\delta$&$-0.61\pi$& $-1.26\pi-0.14\pi$ & $0-2\pi$ \\
& &$[-0.41\pi]$& $[-1.11\pi-0.24\pi]$ & $[0-2\pi]$ \\
\end{tabular}
\end{ruledtabular}
\end{table}
\end{center}

In the same spirit, the large value of $|U_{e3}|$ has implications for LFV in the 
$e\tau$ sector. The condition for $(m_\nu  m_\nu^\dagger)_{e\tau} = 0$
gives the following result: 
{\footnotesize
\bea \label{eq:13zero}
|U_{e3}|_{(m_\nu  m_\nu^\dagger)_{e\tau}=0}&=& \frac{1}{2}  \frac{R \, \sin 2 \theta_{12}  
\tan \theta_{23} }{1 \mp R  \sin^2 \theta_{12}} 
\simeq \frac{1}{2}  R  \sin 2 \theta_{12}  
\tan \theta_{23} \nonumber\\
&=& \left\{ \baz 
0.0146_{-0.003\,(0.006)}^{+0.004\,(0.010)}  & \mbox{normal} \, , \\[.2cm]
0.0153_{-0.003\,(0.006)}^{+0.003\,(0.009)}  & \mbox{inverted}  .
\ea \right.
\eea}
Similar to $(m_\nu  m_\nu^\dagger)_{e\mu}$ one can evaluate the 
right-hand side of Eq.~(\ref{eq:13zero}), giving similar numbers.

LFV processes in the $\tau \mu$ sector also have lower
limits, since the relevant flavor quantity $(m_\nu 
m_\nu^\dagger)_{\mu\tau}$ cannot vanish. This was true even before the recent results on 
$U_{e3}$. At leading order, one finds
\beq\label{eq:23}
|(m_\nu  m_\nu^\dagger)_{\mu\tau}| \simeq 
\frac{1}{2} \Delta m^2_{atm} \, \sin 2 \theta_{23} \left(1 - R \, \cos^2 \theta_{12}
\right) , 
\eeq
which is always non-zero. The order of magnitude 
of $(m_\nu m_\nu^\dagger)_{\mu\tau}$ 
is always larger than the one of $(m_\nu  m_\nu^\dagger)_{e\mu}$: 
\beq\label{eq:12by23}
\frac{|(m_\nu  m_\nu^\dagger)_{e\mu}|^2}{|(m_\nu
\,m_\nu^\dagger)_{\mu\tau}|^2} \simeq 
\frac{|U_{e3}|^2}{\cos^2 \theta_{23}} + 2 |U_{e3}|  \cos \delta \, 
\frac{\sin 2 \theta_{12}}{\sin 2 \theta_{23}} \, R . 
\eeq
We will continue with a study focusing on the decay $\mu \to e
\gamma$ in the type II seesaw, leaving a more detailed study of other
decays and other scenarios for a future study.  In general, however, 
the necessary existence of LFV in the $e\mu$ (and $e\tau $ sector)
adds to the known existence of  LFV in the $\tau \mu$ sector, and
guarantees the presence of all three 
channels.

\section{\label{sec:main}Non-vanishing Branching Ratios: 
Example of the Higgs Triplet}
As mentioned before, we focus here on the type II or triplet seesaw mechanism. 
In this framework neutrino masses are generated by interactions of
lepton doublets $L_\alpha$, with $\alpha = e, \mu, \tau$, with a weak
triplet, hypercharge 2 scalar:
\bea \label{eq:L}
{\cal L} &=& h_{\alpha\beta} \, \overline{L_\alpha^c } i\tau_2 \, \Delta \,
L_\beta + H.c.,~\text{ where }\nonumber\\
\Delta &=& \left(\begin{array}{cc}
H^+ /\sqrt{2} & H^{++} \\ 
H^0 & -H^+ /\sqrt{2}
\end{array}\right).
\eea
Upon acquiring a vacuum expectation value (VEV) $\langle H^0 \rangle =
v_\Delta/\sqrt{2}$, the neutrino mass matrix for light Majorana neutrinos
is 
\beq\label{eq:mnu}
(m_\nu)_{\alpha\beta} = \sqrt{2} \, v_\Delta \, h_{\alpha\beta}  ,
\eeq 
where $h_{\alpha\beta}$ are the neutrino Yukawa couplings.
The interesting and potentially substantiate part of this mechanism is
that the members of the triplet induce 
LFV with couplings given in terms of Eqs.~(\ref{eq:L}) and (\ref{eq:mnu}),
i.e.~in terms of in principle measurable parameters \cite{prev_LFV}. These parameters,
together with the masses of the triplet members which are in principle
accessible at colliders \cite{prev_LHC}, allow for a scenario that is fully
determinable and makes definite predictions for LFV.

Let us recapitulate the well-known formulas for the branching
ratios \cite{prev_LFV}. For $\mu \to e \gamma$ one has 
\beq \label{eq:mueg}
{\rm Br}(\mu \to e \gamma) = \frac{27  \alpha}{256  \pi  G_F^2
 M_{H^{\pm\pm}}^4}  \frac{|(m_\nu
m_\nu^\dagger)_{e\mu}|^2}{v_\Delta^4} {\rm{Br}}(\mu \to e\bar{\nu}\nu) , 
\eeq
with $M_{H^{\pm\pm}}$ as the triplet mass and 
Br$(\mu \to e\bar{\nu}\nu) \simeq 100 \%$. 
The branching ratio for $\tau \to e \gamma$ is given by
\beq \label{eq:teg} 
{\rm Br}(\tau \to e \gamma) = \frac{27  \alpha}{256  \pi  G_F^2
M^4_{H^{\pm\pm}}}  \frac{|(m_\nu  m_\nu^\dagger)_{e\tau}|^2}{v^4_\Delta} 
{\rm Br}(\tau \to e \bar{\nu}\nu), 
\eeq
where ${\rm Br}(\tau \to e \nu\nu) = 17.82 \pm 0.04 \%$
\cite{Nakamura:2010zzi}. The analogous formula for ${\rm Br}(\tau \to
\mu \gamma)$ depends on $(m_\nu  m_\nu^\dagger)_{\mu\tau}$. 
At this stage, combining Eqs.~(\ref{eq:mueg}) and (\ref{eq:teg}), 
we can rewrite Eq.~(\ref{eq:teg}) as
\bea \label{eq:tegapprox} 
{\rm Br}(\tau \to e \gamma) =  0.1782\times\frac{|(m_\nu  m_\nu^\dagger)_{e\tau}|^2}
{|(m_\nu  m_\nu^\dagger)_{e\mu}|^2} 
~{\rm Br}(\mu \to e \gamma)\,. 
\eea
In general, as stated earlier, Br$(\mu \to e \gamma)$ and Br$(\tau \to e
\gamma)$ are of the same order of magnitude since $(m_\nu 
m_\nu^\dagger)_{e\mu} \sim (m_\nu m_\nu^\dagger)_{e\tau}$ due to 
the approximate $\mu$--$\tau$ symmetry of lepton mixing.  
The current limit on Br$(\tau \to e
\gamma)$ is $3.3 \times 10^{-8}$ \cite{Nakamura:2010zzi}, with a potential
improvement to $3.0 \times 10^{-9}$ in the SuperB facility
\cite{future-LFV}, still being way below the current $\mu \to e
\gamma$ limit. Recall that this was recently improved to \cite{Adam:2011ch}
\beq
{\rm Br}(\mu \to e \gamma) < 2.4 \times 10^{-12} \, ,
\eeq
and future limits to values down to $10^{-13}$ are foreseen \cite{Hoecker:2012nu}. 
Exact $\mu-\tau$ symmetry would result in 
$\frac{|(m_\nu  m_\nu^\dagger)_{e\tau}|^2}{|(m_\nu
m_\nu^\dagger)_{e\mu}|^2} = 1$ and thus $\frac{Br(\tau \to e
\gamma)}{Br(\mu \to e \gamma)}\simeq 0.2$. In this case a limit on
Br$(\mu \to e \gamma)<10^{-{12}}$ would correspond to Br$(\tau \to e
\gamma)<10^{-{13}}$, beyond the reach of upcoming
experiments (see Table \ref{constraints}). 
A careful study including the variation of the oscillation
parameters shows that $\frac{Br(\tau \to e
\gamma)}{Br(\mu \to e \gamma)}\simeq 0.15 - 0.21~
({\rm both~for}~1\sigma~{\rm and}~3\sigma)$, 
and hence this
conclusion remains valid.  Thus any evidence of
$\tau \to e \gamma$ in near future experiment will rule out triplet
seesaw models or any model in which $m_\nu m_\nu^\dagger$ governs
LFV.

We should remark that $\mu \to 3 e$ is also a very interesting
process, being mediated at tree level. 
The branching ratio for $\mu \to 3e$ is given by
{\footnotesize
\beq \label{eq:mueee}
{\rm Br}(\mu \to 3e) = \frac{1}{16 G_F^2  M^4_{H^{\pm\pm}}} 
\frac{|(m_\nu)_{\mu e}|^2 |(m_\nu)_{ee}|^2}{v_\Delta^4} {\rm{Br}}(\mu \to e\bar{\nu}\nu).
\eeq}
Unlike $\mu \to e \gamma$,
$\tau \to e \gamma$ or $\tau \to \mu \gamma$, the process $\mu \to 3 e$
can yield an experimentally inaccessible branching ratio even with recent 
$\theta_{13}$ value and low triplet masses, namely when the $ee$ or
$e\mu$ elements of the Majorana neutrino mass matrix vanish. 
In this case, one-loop diagrams can provide the dominant contribution,
depending on $(m_\nu m^\dagger_\nu)_{e\mu}$, the same flavour quantity
that governs $\mu \to e \gamma$. Assuming that the decay is generated
by $e^+ e^-$ pair creation from a virtual photon, the following ratio 
of branching ratio is found: 
\begin{equation} \label{eq:ratio}
\frac{{\rm Br}(\mu \to 3e)}{{\rm Br}(\mu \to e\gamma)} =
\frac{\alpha_{em}}{3\pi} \left[ \log \frac{m_\mu^2}{m_e^2} -
\frac{11}{4} \right] \simeq 1.5 \times 10^{-3} \, . 
\end{equation}
Thus, Br$(\mu \to e \gamma)\sim$ $10^{-12}$
implies Br$(\mu \to 3 e)\sim$ $10^{-15}$. This illustrates the
importance of experiments focusing on dramatic improvement of limits
on $\mu \to 3e$. We note that two proposals are under discussion,
which aim to go down to $10^{-16}$, one at PSI and one at the MuSIC
facility in Osaka. Our finding applies to those possible
experiments. However, there may be cancellations of this
loop-suppressed $m_\nu m_\nu^\dagger$ contribution with a small
tree-level contribution, so that lower limits on Br($\mu \to 3e$) are
not as straightforward as the ones for $\mu \to e \gamma$. 
Since in addition the projects on $\mu \to 3e$ are not as advanced as the
other LFV search experiments, we will not discuss this issue further,
leaving it for further study. 

A related remark concerns the supersymmetric type II seesaw case \cite{Rossi:2002zb},
in which a super-heavy supersymmetric triplet exists at the scale of $B-L$
breaking, and universal boundary conditions for the slepton masses, 
trilinear couplings and gaugino masses are present. Renormalization
group evolution from the Grand Unified scale to the triplet mass generates LFV that
depends solely 
on $m_\nu  m_\nu^\dagger$. Therefore, $\mu \to 3 e$ depends now also on 
$(m_\nu  m_\nu^\dagger)_{e\mu}$, i.e.~on the same quantity as the
branching ratio of $\mu \to e \gamma$. The relative factor between the
two branching ratios is the same as in Eq.~(\ref{eq:ratio}). In this case, the decay $\mu
\to 3 e$ is guaranteed to happen. To quantify the order of magnitude of the LFV
rates one would need to specify the various supersymmetry parameters, which we
will not do here.

Finally the $\mu$ to $e$ conversion rate in nuclei is given by \cite{conv}:
\begin{widetext}

\bea \label{eq:mueconversion} 
{\rm R}(\mu N\to e N^*) = \frac{\alpha^5 m^5_\mu Z^4_{\rm{eff}} 
Z |F(q)|^2}{16 \pi^4 M^4_{H^{\pm\pm}} v^4_\Delta \Gamma_{\rm{capt}}}
\left|\sum_{k=e,\mu,\tau}\frac{(m^\dagger_\nu)_{ek}(m_\nu)_{k\mu} F(r,s_k)}{3}
- \frac{3(m^\dagger_\nu m_\nu)_{e\mu}}{8} \right|^2,
\eea
where 
\bea \label{eq:muefrsk} 
F(r,s_k) = {\rm{ln}}{s_k} + \frac{4s_k}{r}
+\left(1-\frac{2s_k}{r}\right)\sqrt{\left(1+\frac{4s_k}{r}\right)}
~{\rm{ln}}\frac{\sqrt{1+\frac{4s_k}{r}} + 1}{\sqrt{1+\frac{4s_k}{r}} - 1} .
\eea
 
\end{widetext}
with $r = -\frac{q^2}{M^2_{H^{\pm\pm}}}$, $s_k = \frac{m^2_k}{M^2_{H^{\pm\pm}}}$, 
$k=e,\mu,\tau$. For $\mu N \to e N^*$ in different nuclei corresponding values of
$Z_{\rm{eff}},\Gamma_{\rm{capt}},F(q^2\simeq -m^2_\mu)$ can be obtained from 
Ref.~\cite{Kitano:2002mt}. 
The best current limit on the $\mu-e$ conversion 
ratio R($\mu \to e$) is $7\times10^{-13}$ for $^{197}_{79}\!$Au \cite{Nakamura:2010zzi}.
Future experiments ({\tt Mu2e, COMET}, using $^{27}_{13}\rm{Al}$) \cite{mu2e-future} 
are expected to reach a sensitivity  of $2\times10^{-17}$ in the near future. 
In the far future using $^{48}_{22}\rm{Ti}$, the ratio is
expected to be probed down to values of $10^{-18}$
\cite{Barlow:2011zz}. 
As obvious from Eq.~(\ref{eq:mueconversion}), there are two
contributions to the process, and it turns out that setting a lower
limit on the rate of $\mu-e$ conversion is not possible, even with
large $|U_{e3}|$. While the second contribution in ${\rm R}(\mu N\to e
N^*)$ is the same expression as in $\mu \to e\gamma$ and has a lower
limit, it can be cancelled by the more complicated first term, which
depends in a complicated way on the individual neutrino masses and
Majorana phases. In fact, the rate of $\mu-e$ conversion 
under certain assumption, can vanish
for certain parameter values, as recently shown in
Ref.~\cite{Dinh:2012bp}. We will therefore not study this process
anymore and will rather focus on the minimal Br($\mu \to e \gamma$) as
implied by recent data on $U_{e3}$.

\section{\label{sec:num} Results of numerical analysis}
Our observation is here that the large observed value of $|U_{e3}|$ implies that the branching
ratio of the decay $\mu \to e \gamma$ cannot vanish, and hence a lower
limit on its branching ratio arises. We quantify this finding now as a function of the triplet
VEV $v_\Delta$ and the triplet mass 
$M_{H^{\pm\pm}}$. When evaluating the minimal (and maximal) value of
$\mu \to e \gamma$, we vary the neutrino oscillation parameters within
the ranges given in Table \ref{tab:osc}; 
their $1\sigma$ and $3\sigma$ ranges are from
Ref.~\cite{Schwetz:2011zk}, 
and for $\theta_{13}$ we have considered the $1\sigma$ and $3\sigma$
ranges from Daya Bay~\cite{An:2012eh}. The three CP phases were also varied in their
allowed ranges.  We took the current constraints on a large number of
LFV processes into account, which are listed in Table
\ref{constraints}. Moreover, we also considered the case of when all
processes obey limits obtainable in future experiments; most of the
future limits have been taken from Ref.~\cite{future-LFV}.  

We have studied the variation of the lowest possible branching ratio
for $\mu \to e \gamma$ with the triplet mass $M_{H^{\pm\pm}}$ for four
different triplet VEVs, $v_\Delta = 0.5$ eV, $1.0$ eV,  $5.0$ eV and
$10.0$ eV.  In the course of
investigation we have also considered the impact of the absolute 
neutrino  
mass scale ($m_1$ for normal hierarchy and $m_3$ for inverted hierarchy)
for three different values, namely $0.003~{\rm{eV}}, 0.05~{\rm{eV}}$ and $0.2~{\rm{eV}}$. 
These values are chosen in a fashion that not only they covered the pure normal
and inverted hierarchical ($m_{1(3)}=0.003$ eV) scenarios, 
but also the quasi-degenerate and intermediate cases. While 
the branching ratio of $\mu \to e \gamma$ does not depend on those
masses, as well as on the Majorana phases, there is an indirect
influence from the limits on the other LFV processes.   

\begin{center}
\centering
\begin{table}[t]
\footnotesize
\caption{\label{constraints}
List of constraints on different lepton flavour violating decays
that have been used in our numerical analysis.
}
\begin{ruledtabular}
\begin{tabular}{cc c c  }
& Process & \multicolumn{2}{c}{Constraints}  \\ \hline
&  & PRESENT &  FUTURE \\ \hline
& Br$(\tau \to e \bar{e} e)$ & $ 2.7\times10^{-{08}}$
\cite{Nakamura:2010zzi}& $1.0\times10^{-{09}}$ \cite{future-LFV}\\ 
& Br$(\tau \to e \bar{e} \mu)$& $1.8\times10^{-{08}}$ \cite{Nakamura:2010zzi}&  $1.0\times10^{-{09}}$ \cite{future-LFV}\\ 
& Br$(\tau \to e \bar{\mu} \mu)$ & $ 1.7\times10^{-{08}}$ \cite{Nakamura:2010zzi}&  $1.0\times10^{-{09}}$ \cite{future-LFV}\\ 
& Br$(\tau \to \mu \bar{\mu} \mu)$&$  2.1\times10^{-{08}}$ \cite{Nakamura:2010zzi}&  $1.0\times10^{-{09}}$ \cite{future-LFV}\\ 
& Br$(\tau \to \mu \bar{e} \mu)$ & $ 1.8\times10^{-{08}}$ \cite{Nakamura:2010zzi}&  $1.0\times10^{-{09}}$  \cite{future-LFV}\\ 
& Br$(\tau \to \mu \bar{e} e)$& $ 1.5\times10^{-{08}}$ \cite{Nakamura:2010zzi}&  $1.0\times10^{-{09}}$ \cite{future-LFV}\\ 
& Br$(\tau \to \mu \gamma)$& $4.4\times10^{-{08}}$ \cite{Nakamura:2010zzi}& 
$2.0\times10^{-{09}}$ \cite{future-LFV}\\ 
& Br$(\tau \to e \gamma)$& $3.3\times10^{-{08}}$ \cite{Nakamura:2010zzi}& 
$3.0\times10^{-{09}}$ \cite{future-LFV}\\
& Br$(\mu \to e \gamma)$& $2.4\times10^{-{12}}$  \cite{Adam:2011ch}& 
 $1.0\times10^{-{13}}$ \cite{future-LFV}\\
& Br$(\mu \to e \bar{e} e)$& $ 1.0\times10^{-{12}}$ \cite{Nakamura:2010zzi}&  $1.0\times10^{-{13}}$ \cite{future-LFV}\\
& R$(\mu \to e)$& $ 7.0\times10^{-{13}}$ (${_{97}^{197}\!{\rm{Au}}}$)
\cite{Nakamura:2010zzi}&  
 $2.0\times10^{-{17}}$ (${^{27}_{13}{\rm{Al}}}$) \cite{future-LFV,mu2e-future}\\
\end{tabular}
 
\end{ruledtabular}
\end{table}
 
\end{center}

\begin{center}
\begin{figure*}[t]
\includegraphics[width=17.55cm,height=13.50cm,angle=0]{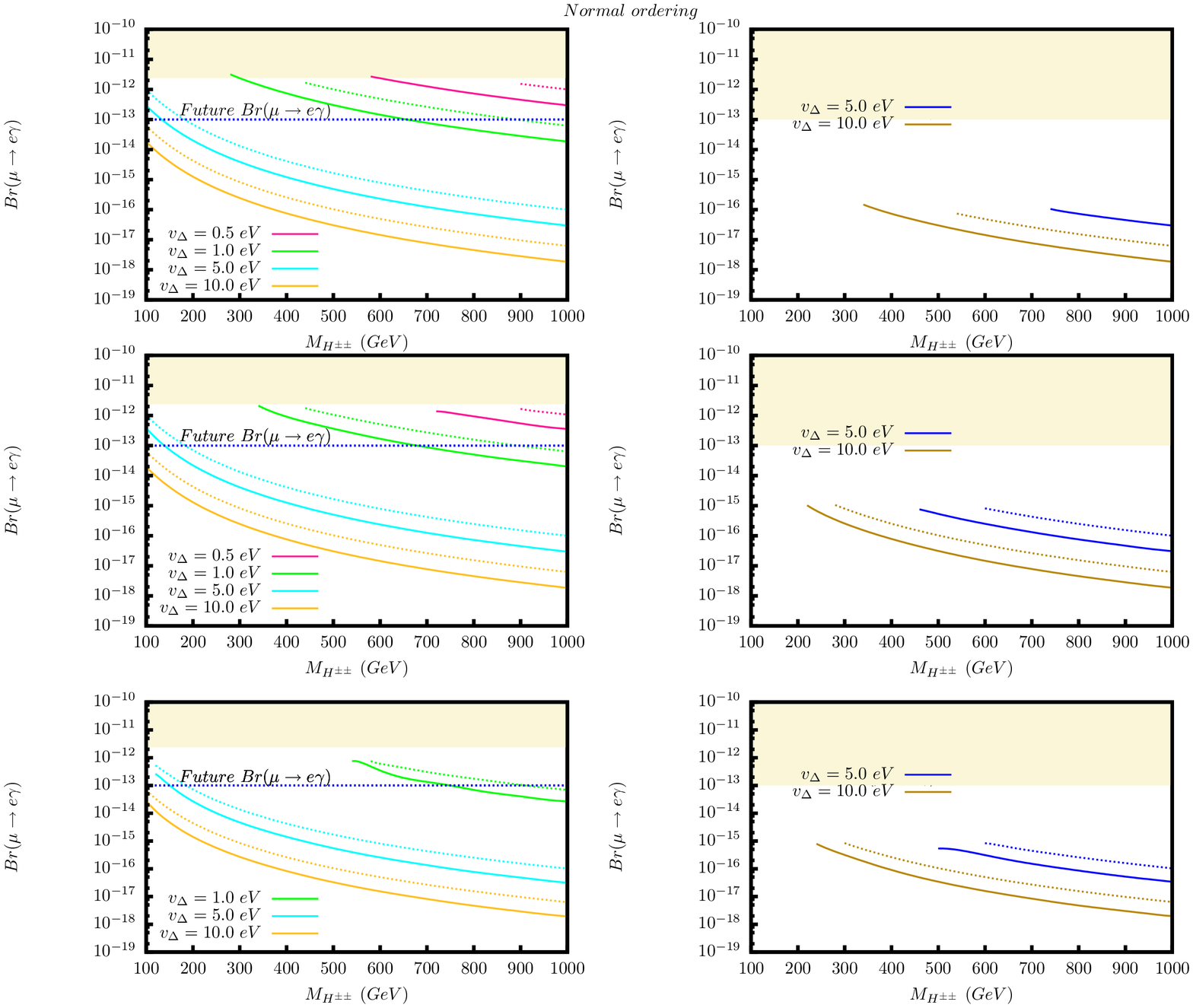}
\caption{ \small {Plots showing the variation of the lowest possible ${\rm Br}(\mu \to e \gamma)$
vs.~$M_{H^{\pm\pm}}$ with different values of $v_\Delta$ for the normal
neutrino mass ordering. The left plots are considering the present
constraints on different LFV processes and the right ones are with the future constraints.
Plots in the upper row are with the lightest neutrino mass $m_1=0.003$ eV, 
the middle row is for $m_1=0.05$ eV and the lower row is for $m_1=0.2$~eV. 
The solid (dotted) line corresponds to the $3\sigma$ ($1\sigma$) range
of the oscillation parameters. 
The colored (dark) band corresponds to the exclusion region as
suggested by present and future experimental bounds. 
All constraints are listed in Table \ref{constraints}. The
corresponding plots for the inverted ordering look basically identical. }} 
\label{f:mu2egvsmdel-vdel-NH}
\end{figure*}
\end{center}

It is well understood from
Eqs.~(\ref{eq:mueg}),~(\ref{eq:teg}),~(\ref{eq:mueee}) and
(\ref{eq:mueconversion}) that the branching ratios   
will decrease for larger $M_{H^{\pm\pm}}$ and $v_\Delta$. 
Consequently, if we ask that the stronger future constraints are
obeyed, larger $M_{H^{\pm\pm}}$ and $v_\Delta$ are more favorable. 
Further, with light $v_\Delta$, larger triplet masses are favorable. 
Of course, for sufficiently large values of triplet mass and VEV, some of these 
branching ratios will be inaccessible to the ongoing and even to the future experiments.
In addition there may arise situations when some of the processes
remain unobserved while others have been seen. Such more 
complicated situations will be discussed elsewhere.

\begin{center}
\begin{figure*}[t]
\includegraphics[width=17.75cm,height=13.50cm,angle=0]{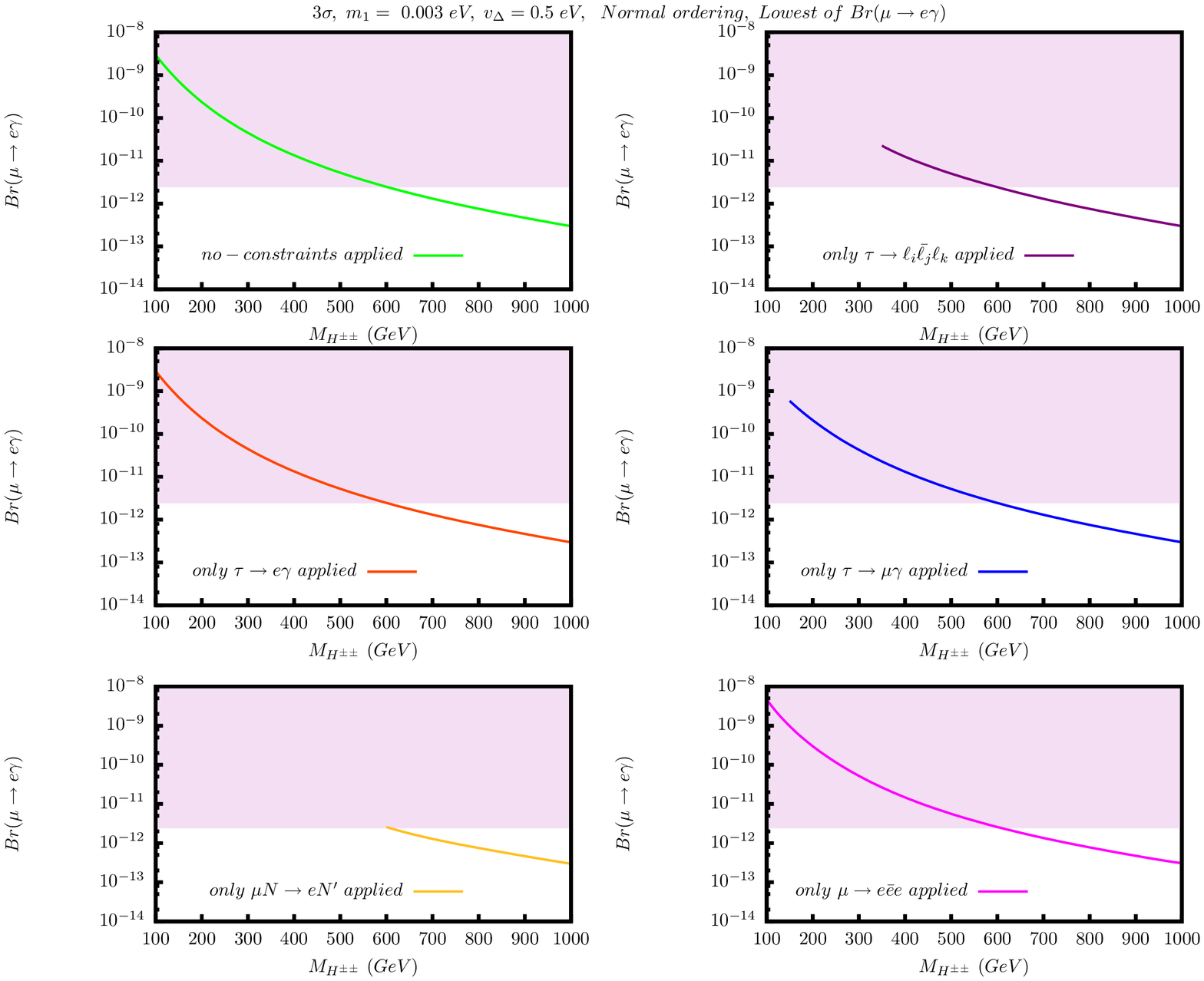}
\caption{ \small {Plots showing the effects of the constraints on other LFV  
processes on the minimal branching ratio of $\mu \to e \gamma$. 
The colored (dark) band corresponds to the exclusion region as
suggested by the present experimental bound.}} 
\label{f:mu2egvsmdel-constraints-effect-NH}
\end{figure*}
\end{center}


Varying over the oscillation parameters, one expects very similar
behavior for the normal and inverted ordering (there are only tiny
differences because the indirect constraints from other LFV processes
depend on the mass ordering).  
Therefore, we only plot the normal ordering case in
Fig.~\ref{f:mu2egvsmdel-vdel-NH}. As can be seen,  with lighter
$v_\Delta=0.5$ and $1.0$ eV, the region with lighter triplet mass is 
excluded by the other LFV constraints. With the present constraints,
there exists no allowed region for $v_\Delta=0.5$ eV and
$m_{1(3)}=0.2$ eV. 
Obviously with heavier triplet mass $(M_{H^{\pm\pm}}>1$ TeV),
such conclusion no longer remains valid. Nevertheless, scenario with a
very heavy triplet has less appealing collider phenomenology. 
We have noted that throughout all the parameter space $\mu - e$ conversion 
posses the most stringent bounds. 
With the future constraints, exclusion of the
entire region with any values of triplet mass and for $v_\Delta=0.5$ and $1.0$ eV,
is solely due to the very stringent future $\mu - e$ conversion
constraint \cite{mu2e-future}.  As can be seen from
Fig.~\ref{f:mu2egvsmdel-vdel-NH}, pushing the branching ratio of $\mu
\to e \gamma$ down to $10^{-13}$ makes it possible to definitely probe
regions of parameter space of $v_\Delta$ and
$M_{H^{\pm\pm}}$. Examples are if $v_\Delta \ls 5$ eV and
$M_{H^{\pm\pm}} \ls 200 $ GeV, or when $v_\Delta \ls 1$ eV and
$M_{H^{\pm\pm}} \ls 700 $ GeV. We stress again that before the recent
results on large $U_{e3}$ were obtained this was not possible. 
The effects of the constraints of
the other LFV modes on the minimal value of Br$(\mu \to e \gamma)$ can
be seen in Fig.~\ref{f:mu2egvsmdel-constraints-effect-NH}. Two
implications are resulting when one switches on the other LFV limits:
(i) the scale of $M_{H^{\pm\pm}}$ is set to larger values, and (ii) the lower limit on
the branching ratio is increased by a moderate amount.

\section{\label{sec:concl}Summary}
Lepton Flavor Violation (LFV) may be connected directly or indirectly
to neutrino oscillation parameters. In this paper we worked in 
scenarios with presumably the most direct connection, in which the
quantity $m_\nu  m_\nu^\dagger$ is responsible for LFV in the charged
lepton sector. Minimal flavor violation in the lepton sector, as well
as other frameworks and scenarios, has such a feature.  
We noted that recent results 
on the lepton mixing parameter $U_{e3}$ imply that the $(m_\nu
m_\nu^\dagger)_{e\mu}$ cannot vanish. Consequently, lower limits on  
lepton flavor violation arise, and we have quantified this with the
example of $\mu \to e \gamma$ in the type II seesaw mechanism, in
which a Higgs triplet is responsible for neutrino mass. We stress that
many more examples in which our finding applies can be discussed. 

We also shortly discussed processes as $\mu \to 3e$ and $\mu-e$
conversion, where $(m_\nu m_\nu^\dagger)_{e\mu}$ is also of
relevance. However, either the contribution of $(m_\nu
m_\nu^\dagger)_{e\mu}$ is suppressed, or cancellations from other
contributions can occur. Setting lower limits in the same sense as for
$\mu \to e\gamma$ is not possible.

While searches for lepton flavor violation do not need further
motivation, we feel that our observation closes yet another loophole
that would allow LFV to hide, and adds additional interest to study
LFV in the $e\mu$ sector.

\vspace{0.3cm}
\begin{center}
{\bf ACKNOWLEDGMENTS}
\end{center}
W.R. is supported by the ERC under the Starting Grant 
MANITOP and by the DFG in the project RO 2516/4--1. 
P.G. is supported by the Spanish MICINN under grant FPA2009-08958. P.G.
also acknowledges the support of  the MICINN under the Consolider-Ingenio
2010 Programme with grant MultiDark CSD2009-00064, the Community of
Madrid under grant HEPHACOS S2009/ESP-1473, and the European Union
under the Marie Curie-ITN program PITN-GA-2009-237920. We acknowledge
the hospitality provided by the organizers of WHEPP-XII held at
Mahabaleshwar, India where this work was initiated. 


\end{document}